\documentclass[sigconf]{acmart}

\AtBeginDocument{%
  }

\setcopyright{acmcopyright}
\copyrightyear{2018}
\acmYear{2018}
\acmDOI{XXXXXXX.XXXXXXX}

\acmConference[Conference acronym 'XX]{Make sure to enter the correct
  conference title from your rights confirmation emai}{June 03--05,
  2018}{Woodstock, NY}
\acmPrice{15.00}
\acmISBN{978-1-4503-XXXX-X/18/06}

\usepackage{graphicx}
\usepackage{subfigure}
\usepackage{booktabs} 
\usepackage{amsmath}

\usepackage{amssymb}
\usepackage{times}
\usepackage{algorithm}
\usepackage{algorithmic}
\usepackage{xspace}
\usepackage{multirow}
\usepackage{threeparttable}

\newcommand{\hide}[1]{}
\newcommand{\name} {{\textsc{DTInspector}}\xspace}
\newcommand{\yy}[1]{{\color{red} YY: #1}}

\newcommand{\abs} {{\textsc{ABS}}\xspace}

\newcommand{\nc} {{\textsc{NC}}\xspace}

\newcommand{\badnet} {{\textsc{BadNet}}\xspace}
\newcommand{\tronn} {{\textsc{TrojanNN}}\xspace}
\newcommand{\cl} {{\textsc{CL}}\xspace}
\newcommand{\sig} {{\textsc{SIG}}\xspace}
\newcommand{\ulp} {{\textsc{ULP}}\xspace}
\newcommand{\meta} {{\textsc{(MNTD}}\xspace}
\newcommand{\refool} {{\textsc{Refool}}\xspace}
\newcommand{\Ftrojan} {{\textsc{Ftrojan}}\xspace}

\def \cR {\mathcal{R}}
\def \bg {\mathbf{g}}
\def \be {\mathbf{e}}
\def \bbE {\mathbb{E}}
\def \bbR {\mathbb{R}}

\def \cX {\mathcal{X}}
\def \cY {\mathcal{Y}}

\def \cG {\mathcal{G}}
\def \fR {\mathfrak{R}}

\begin{document}

\title{Confidence Matters: Inspecting Backdoors in Deep Neural Networks via Distribution Transfer}

\author{
    Tong Wang$^1$, Yuan Yao$^1$, Feng Xu$^1$
    Miao Xu$^2$, 
    Shengwei An$^3$, 
    Ting Wang$^4$
}
\affiliation{
    \institution{$^1$State Key Laboratory for Novel Software Technology, Nanjing University, China \\
    $^2$The University of Queensland, Australia\\
    $^3$Purdue University, USA\\
    $^4$Pennsylvania State University, USA \country{}}
}
\email{mg20330065@smail.nju.edu.cn, {xf, y.yao}@nju.edu.cn,  miao.xu@uq.edu.au, an93@purdue.edu, inbox.ting@gmail.com}

\renewcommand{\shortauthors}{Wang et al.}

\begin{abstract}
Backdoor attacks have been shown to be a serious security threat against deep learning models, and detecting whether a given model has been backdoored becomes a crucial task. Existing defenses are mainly built upon the observation that the backdoor trigger is usually of small size or affects the activation of only a few neurons. However, the above observations are violated in many cases especially for advanced backdoor attacks, hindering the performance and applicability of the existing defenses.
In this paper, we propose a backdoor defense \name built upon a new observation. That is, an effective backdoor attack usually requires high prediction confidence on the poisoned training samples, so as to ensure that the trained model exhibits the targeted behavior with a high probability. Based on this observation, \name first learns a patch that could change the predictions of most high-confidence data, and then decides the existence of backdoor by checking the ratio of prediction changes after applying the learned patch on the low-confidence data. Extensive evaluations on five backdoor attacks, four datasets, and three advanced attacking types demonstrate the effectiveness of the proposed defense. 
\end{abstract}
\maketitle

\section{Introduction}

Despite showing great potential in many practical applications, deep learning models have been found to be vulnerable to backdoor attacks (a.k.a. trojan attacks)~\cite{gu2017badnets}. For example, by poisoning the training data, a face recognition model can be easily manipulated to predict any person as the specified target person.
Typically, a successful backdoor attack poisons a small set of training data, and renders a trained model satisfying the following two requirements: 1) {\em stealthiness}, the trojaned model does not have significant accuracy degradation on benign inputs; 2) {\em effectiveness}, it classifies a poisoning input (e.g., input stamped with a trigger) to a target label with a high probability. 



Up to date, 
several defenses have been proposed to inspect whether a given model has been trojaned or not. 
Among them, \citet{wang2019neural} propose Neural Cleanse (\nc) which is built upon the intuition that a backdoor trigger is usually of small size, and it is observed to be sensitive to the trigger size, performing poorly when the trigger size is relatively large~\cite{guo2020tabor}.
\citet{liu2019abs} propose \abs which analyzes the neuron activation to detect backdoor, based on the assumption that stimulating one inner neuron of a trojaned model can easily lead to the target label. However, this assumption is too strong making \abs fall short against advanced attacks involving multiple triggers or multiple target labels~\cite{gao2020backdoor}.
Kolouri et al.~\cite{kolouri2020universal} propose \ulp which trains a meta-classifier based on a number of clean and trojaned models. However, the trained meta-classifier may overfit the training data and cannot generalize well to unseen attacks or triggers.

To address the limitations of existing defenses, in this work, we propose a backdoor defense \name from a novel perspective. Specifically, our method is motivated by the observation that an effective backdoor attack usually needs to achieve high prediction confidence on the poisoning data at training stage, so as to ensure high attack success rate (ASR) on the poisoning inputs at inference stage. This observation does not rely on the trigger size or the neuron activation pattern, and thus has the potential to address the above limitations. We provide both theoretical and empirical evidence for this observation.

Making use of the above observation is non-trivial. A straightforward backdoor way is to directly compare the distributions between high-confidence and low-confidence data. However, such comparison is less effective as the distributions between clean samples and poisoning samples are difficult to distinguish, especially when the trigger involves only a small perturbation.
To overcome this issue, we propose a {\em distribution transfer} technique, which is built upon the shortcut nature~\cite{geirhos2020shortcut,li2021anti} of triggers. Specifically, for each label, we first learn a {\em patch} that could change most predictions on a few high-confidence data, and then compute the {\em transfer ratio}, i.e., the ratio of prediction changes, by applying this patch on the low-confidence data. For an infected/target label, the learned patch tends to destroy the trigger, as this is the easiest way (i.e., shortcut) to change the predictions (e.g., back to the original labels); however, the patch becomes less effective in terms of changing the predictions of low-confidence data that does not contain triggers. On the contrary, for a clean label, the learned patch tends to be universal adversarial perturbations~\cite{moosavi2017universal} that are more difficult to learn compared to the shortcut, and thus will also change most predictions of the low-confidence data.

We build \name by employing the above insight, and evaluate it against five basic backdoor attacks and three advanced attacking types on four datasets. The studied advanced attacks~\cite{wang2019neural} include partial attack (poisoning data from a specific label), MTOT attack (multiple triggers for one target label), and MTMT attack (multiple triggers for multiple target labels).
The results show that our defense can accurately detect the trojaned models as well as the infected labels. We also compare \name with three state-of-the-art defenses (i.e., \nc~\cite{wang2019neural}, \abs~\cite{liu2019abs}, and \ulp~\cite{kolouri2020universal}) that aim to detect whether a given model is trojaned. The results show that \name outperforms these competitors: 1) it is less sensitive to the trigger size, 2) it can successfully detect advanced attacks, 3) it can be naturally applied to unseen attacks/triggers.

In summary, the major contributions of this paper include:
\begin{itemize}
\item A new perspective for detecting backdoor attacks which is both theoretically and empirically supported.
\item A new backdoor detection method \name that can handle advanced attacking types.
\item Extensive experimental evaluations showing the effectiveness of \name.
\end{itemize}



\section{Threat Model and Key Observation}




\subsection{Threat Model}

We consider the commonly-studied poisoning-based backdoor attacks~\cite{gu2017badnets,liu2018trojaning,turner2018clean}. That is, the adversary defines a trigger beforehand and then injects it into a few training data,  
but does not have access to the training process and the model.\footnote{This assumption is practical in reality. For example, the user training the model may have collected some data from the Web, and the collected data may contain poisoning samples deliberately published by malicious attackers.}
By corrupting training data, the adversary is able to launch advanced attacks. The adversary can also adaptively tune the poisoning rate to lower down the prediction confidence. We assume that the adversary only attacks a minority of labels.
For the defender, we assume she can access both the training data and the model, and she does not require extra clean data. The goal of the defender is to determine whether the trained model has been implanted a backdoor.
\subsection{Key Observation}
Our key observation is that an effective backdoor attack usually needs to achieve high prediction confidence on the poisoned training data, so as to ensure high ASR.

\noindent{\bf Theoretical Results}.
To verify our observation, we first analyze the relationship between the prediction confidence on poisoned training samples and the ASR on poisoning inputs. Assume the data point $(X,Y)$ is sampled uniformly at random from $\cX \times \cY$, where $\cX = \bbR^d$ and $\cY = [c]$. Take multi-class learning as a typical example, and it aims to optimize the following risk,
\small
\begin{eqnarray*}
\cR(\bg) = \bbE_{(X,Y)\sim p(x,y)}[\ell(\bg(X), \be^Y)]
\end{eqnarray*}
\normalsize
with a minimizer $\bg^*$: $\cX\rightarrow \bbR^c$. Here $\be^Y$ is the standard canonical vector with the $Y$-th entry to be one and all others zero. $\ell$ is often defined to be the cross-entropy loss. 

Given a backdoor attack which poisons some training data in different labels to the target label $t$, and using $a_z$ to denote the poisoning rate in label $z$, the above risk becomes 
\small
\begin{eqnarray*}
\cR_a(\bg) &=& \sum_{z \in [c]} a_z \cdot p(Y=z)\bbE_{X\sim p(x|y = z)}[\ell(\bg(X), \be^{t})] + \\
&& \sum_{z \in [c]} (1-a_z)p(Y=z)\bbE_{X\sim p(x|y = z)}[\ell(\bg(X), \be^{z})]
\end{eqnarray*}
\normalsize
with a minimizer denoted as $\bg_a^*$. Here, we assume that the trigger involves very small perturbations and can be approximately ignored in the above equation.
Based on the above two equations, we have the following theorem.

\begin{theorem}\label{Th:1}
For an effective backdoor attack that does not change the prediction of clean data, with a probability at least $1-\delta$
\small
\begin{eqnarray*}
\cR(\bg_a^*) - \cR(\bg^*) \le 4\fR_n(\ell\circ \cG) + 2M\sqrt{\frac{\log(2/\delta)}{2n}}+\\
2 \sum_{z \in [c]} a_z \sum_{x_i\in D_z}|\ell(\bg^*(x_i), \be^z) - \ell(\bg_a^*(x_i), \be^{t})|
\end{eqnarray*}
\normalsize
where $\fR_n(\cdot)$ is the Rademacher Complexity of a function class for sample size $n$, $M$ is the upper bound of the loss function $\ell$, and $D_z$ is the subset of poisoned training data with label $z$. 
\end{theorem}
\textsc{Proof}. See the appendix.

{\em Remarks}. $\cR(\bg^*)$ and $\cR(\bg_a^*)$ are the lowest risks we could get on the given data distribution and function class. Higher $\cR(\bg_a^*) - \cR(\bg^*)$ indicates that the trojaned classifier $\bg_a^*$ incurs a higher risk than the clean classifier $\bg^*$. In other words, higher $\cR(\bg_a^*) - \cR(\bg^*)$ indicates the possibility of higher ASR. 
In the above theorem, $\ell(\bg^*(x_i), \be^z)$ stands for the empirical loss of data sample $x_i$ before poisoning, and $\ell(\bg_a^*(x_i), \be^t)$ stands for its empirical loss after poisoning.
Therefore, the above theorem gives the insight that, when the prediction confidence of data samples before poisoning is close to that after poisoning (i.e., $\ell(\bg^*(x_i), \be^z)$ is close to $\ell(\bg_a^*(x_i), \be^t)$), and the number of poisoning data $\sum_{z} a_z |D_z|$ is small, it is less likely to have an effective backdoor attack with high ASR. 

\begin{figure}[t]
\centering
 \subfigure[Prediction confidence and BA/ASR vs. poisoning rate.]{\label{fig:fig_distribution2}
   \includegraphics[width=0.65\columnwidth]{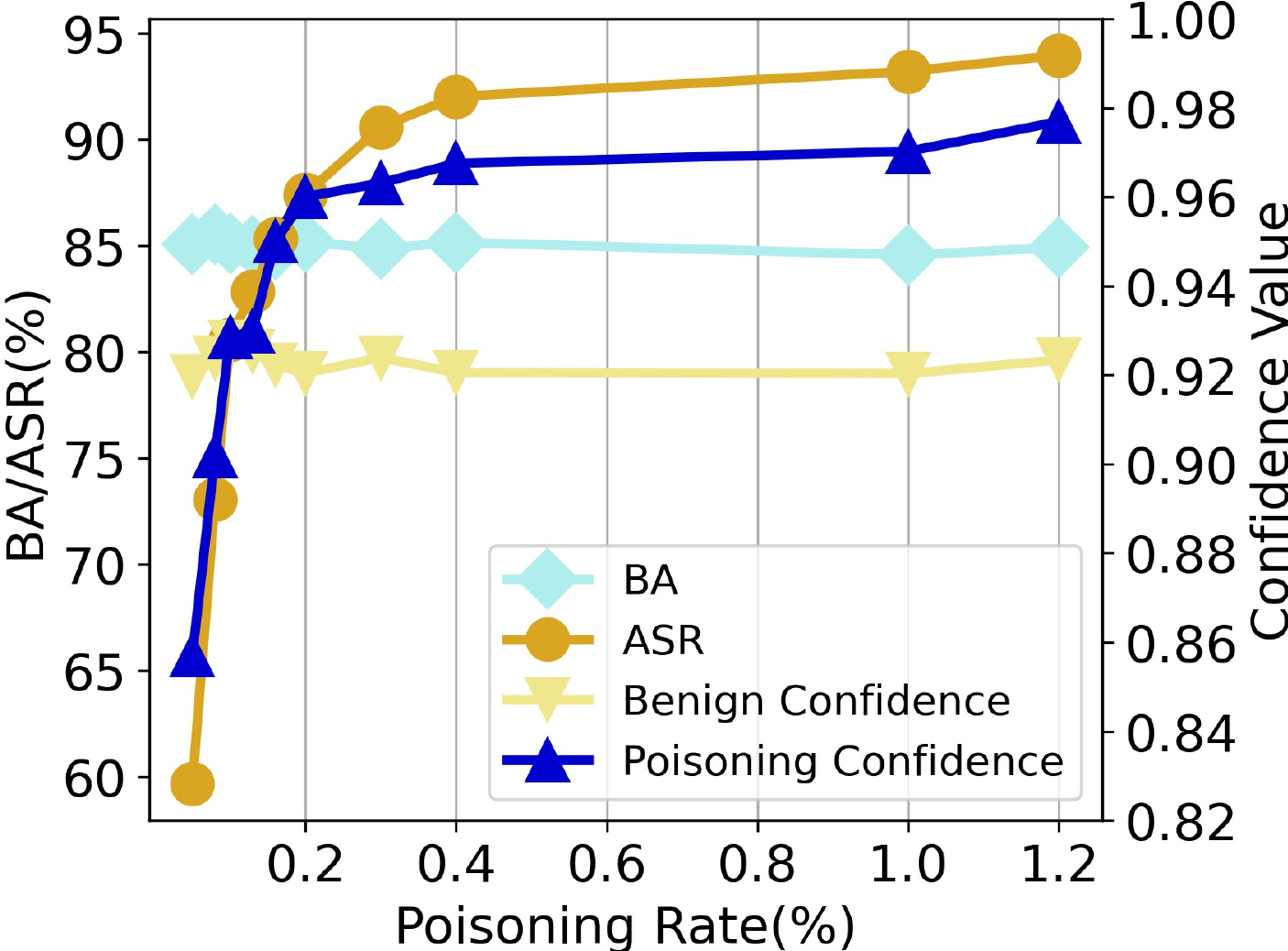}}
 \subfigure[Prediction confidences of poisoning and clean data.]{\label{fig:fig_distribution1}
   \includegraphics[width=0.98\columnwidth]{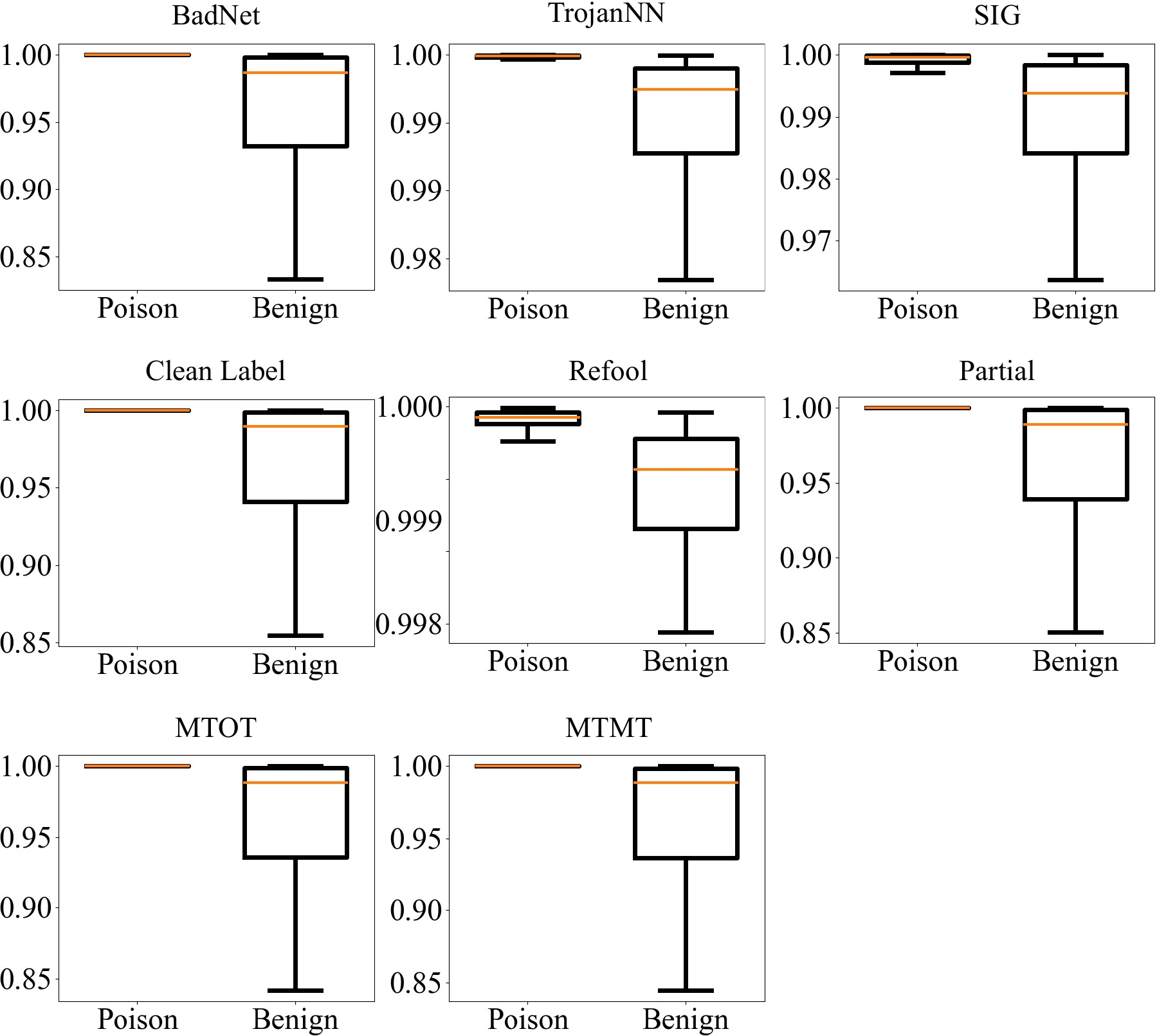}}
\caption{The empirical results for our observation: (a) as the poisoning rate increases, the average prediction confidence of poisoning data becomes significantly higher than that of the clean data; (b) the prediction confidences of most poisoning data are significantly higher than that of the clean data.} 
\label{fig:fig_distribution}
\end{figure}

\noindent{\bf Empirical Results}.
To further verify our observation, we provide some empirical evidence here.
Specifically, we show that, when we slightly increase the poisoning rate to ensure a highly effective backdoor attack, the confidence of poisoning data becomes significantly higher than that of clean data. 
We conduct experiments with various backdoor attacks including \badnet~\cite{gu2017badnets}, \tronn~\cite{liu2018trojaning}, \cl~\cite{turner2018clean}, \sig~\cite{barni2019new}, and \refool~\cite{liu2020reflection}. The results are shown in Figure~\ref{fig:fig_distribution}.

In Figure~\ref{fig:fig_distribution2}, we vary the poisoning rate and show the average results of benign accuracy (BA) on clean inputs, ASR on poisoning inputs, and prediction confidences of poisoned training data and clean training data in the target label. For brevity, we show the results of \badnet on the CIFAR10 dataset here, and similar results are observed on the other attacks which can be found in Appendix~\ref{App:observation}. We can first observe that BA and prediction confidence of clean data are relatively stable indicating the stealthiness of the backdoor attack. Second, as the poisoning rate increases, along with the increase of ASR, the average prediction confidence of poisoning data also becomes significantly higher than that of clean data. For example, when ASR increases up to 85\%, the average prediction confidence of poisoning data is 3.3\% higher than that of clean data (0.95 vs. 0.92).

In Figure~\ref{fig:fig_distribution1}, we further show the distributions of prediction confidences of poisoning data and clean data in the target label. To ensure the effectiveness of each backdoor attack, in this experiment, we gradually increase the poisoning rate by 0.1\% each step until the ASR is above 95\%.  
As we can see, the confidences of most poisoning samples are significantly higher than those of clean samples, for all the five backdoor attacks and three advanced attacking types (i.e., partial attack, MTOT attack, and MTMT attack).

Overall, the above theoretical and empirical results show that it is less likely for a backdoor attack to be highly effective, while keeping the prediction confidences of poisoning data to be close to that of clean data.

\section{The Proposed Defense}

Based on the above observation, a straightforward way is to directly compute the distribution similarity between high- and low-confidence samples. However, as empirically shown by our experimental results in Section~\ref{sec:exp1}, such a method would be less effective since the poisoning perturbation is usually very small.
Therefore, we propose a distribution transfer technique to magnify their difference via using the shortcut nature of triggers.

\subsection{Overview}
 

\begin{figure}[t]
\centering
\includegraphics[width=0.999\linewidth]{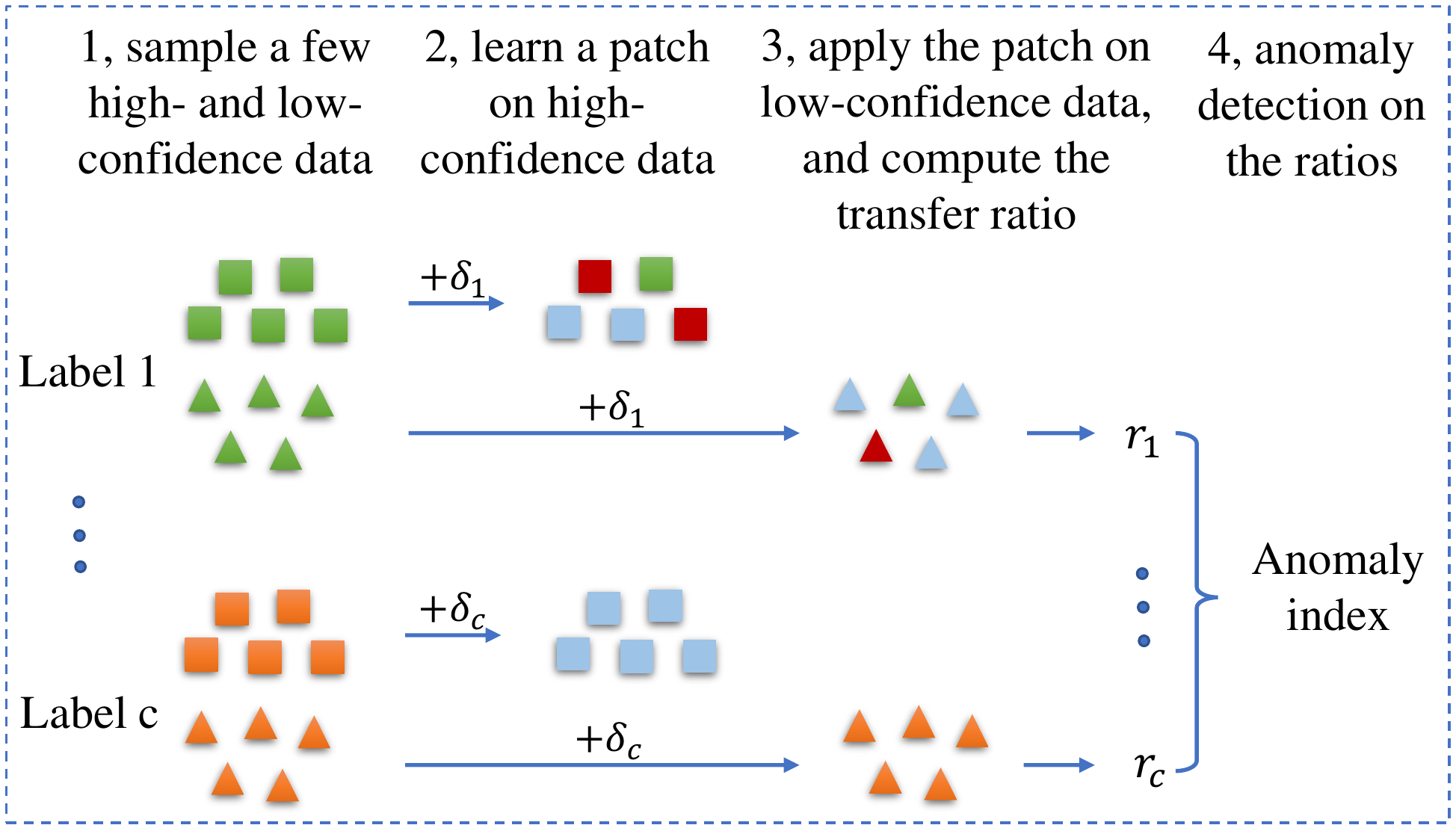}
\vspace{-2ex}
\caption{The overview of \name. Squares and triangles represent high-confidence and low-confidence samples, respectively. Colors indicate the prediction labels. The key insight is that the transfer ratio of infected labels (e.g., Label $c$) would be significantly lower than that of clean labels (e.g., Label 1).}
\label{fig:fig_framework}
\end{figure}

The overview of \name is shown in Figure~\ref{fig:fig_framework}.
Specifically, \name first sorts the training data by confidence for each prediction label $z$, and selects top-$K$ samples from the high-confidence area and bottom-$K$ samples from the low-confidence area. Second, \name learns a small patch $\delta$ using only the high-confidence samples, with the goal of changing most of their predictions to any other labels but the current label $z$. Third, \name applies the learned patch $\delta$ to the low-confidence data, and records the {\em transfer ratio} $r$ meaning the ratio of samples whose prediction labels changed after applying patch $\delta$. Finally, \name performs anomaly detection on all the transfer ratios to spot the trojaned model and the infected label. If the transfer ratios of certain labels are too small, the model is suspected of being attacked.

\hide{Our assumption allows multiple labels to be attacked, but most of them should be clean so that anomaly detection can perform correctly.} 

The key insight of \name is as follows. For a clean label, the learned patch from the high-confidence data can be seen as a universal adversarial perturbation~\cite{moosavi2017universal}, and thus the transfer ratio of low-confidence data will also be high (e.g., Label 1 in Figure~\ref{fig:fig_framework}).
In contrast, for an infected label, the learned patch tends to destroy the trigger due to the shortcut nature of triggers. That is, a trigger is a shortcut that leads the input to the target label~\cite{geirhos2020shortcut,li2021anti}, and destroying this shortcut is the easiest way to change the predictions of the high-confidence data (e.g., back to the original labels). This patch will not largely affect the predictions of low-confidence data that does not contain triggers, making the transfer ratio significantly lower for low-confidence samples in the infected label (e.g., Label $c$ in Figure~\ref{fig:fig_framework}).

Note that \name is also useful to mitigate the effect of backdoor by deleting the potentially poisoned training data (e.g., we superimpose the learned patch to each sample from the infected label and remove the sample if it is classified to a label different from the original label) and then retraining the model. We find that such a strategy can significantly lowering down the ASR, and meanwhile preserves the accuracy in most cases (see Appendix~\ref{App:mitigation} for details).

\subsection{Patch Learning}

For patch learning, 
we define a patch as a tuple $(M, P)$, where $M$ is the 2D mask and $P$ contains the 3D patch pixels. 
We then learn the patch that can lead the high-confidence data to predict a different label. That is, for a certain label $z$ and high-confidence data $D_h$, we aim to optimize the following objective function,
\small
\begin{eqnarray}\label{eq:reg}
    \max_{M, P} & & \sum_{x \in D_h} f(\bg(x'), \be^{z}) - \lambda ||M||_1, \quad  x \in D_{h}, \nonumber\\
    & & \mathrm{s.t.} \quad x' = (1 - M) \odot x + M \odot P,
\end{eqnarray}
\normalsize
where $M$ controls the area to be patched, $P$ controls the value of the patch, and  $\odot$ is element-wise multiplication. $M_{ij}$ ranges from 0 to 1 indicating how much of the original pixel values will be retained or refilled with $P$. For function $f$, since we aim to lead the prediction to any other label but the current $z$, we use the mean square error on the $z$-th dimension. In other words, the above objective encourages the $z$-th dimension of $\bg(x')$ to lean towards zero. $\lambda$ is used to balance the importance of $M$'s $L_1$ norm. Smaller $\lambda$ tends to yield larger patches. $\lambda$ will be dynamically adjusted during the optimization process in order to keep a relatively high transfer ratio (e.g., no less than 95\%).

\begin{figure}[t]
\centering
\includegraphics[width=0.8\columnwidth]{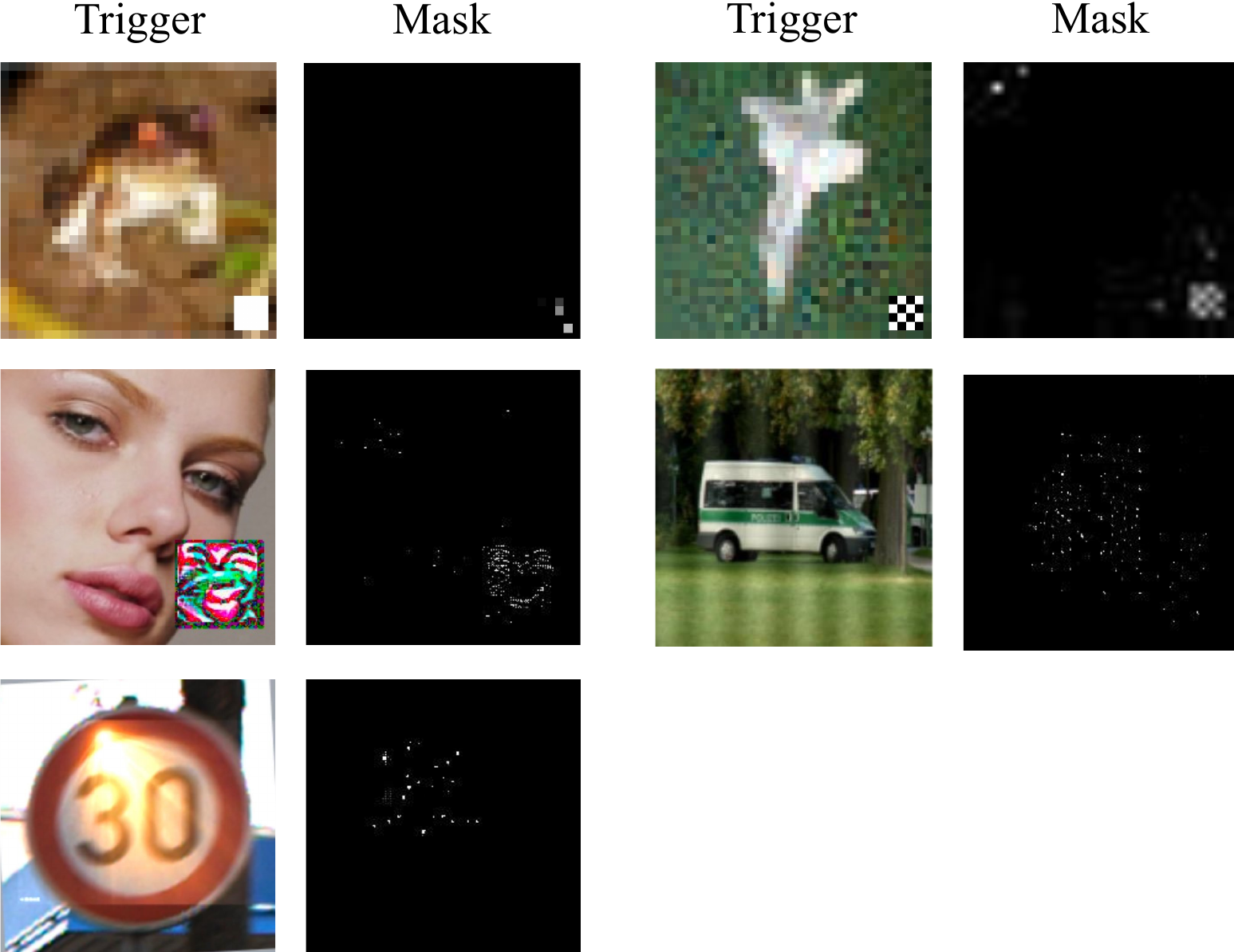}
\caption{The poisoning images and the learned patches by \name. The five examples correspond to five attacks, i.e., \badnet~\cite{gu2017badnets}, \tronn~\cite{liu2018trojaning}, \cl~\cite{turner2018clean}, \sig~\cite{barni2019new}, and \refool~\cite{liu2020reflection}, respectively.}
\label{fig:fig_mask}
\end{figure}

Some examples of the triggered images and the corresponding learned patches (masks) are shown in Figure~\ref{fig:fig_mask}. We can observe that the learned patches can identify the trigger areas. Note that the learned patch does not necessarily need to erase the entire trigger, but only needs to destroy the key pixels, so as to make the predictions of poisoning images from the target label back to their original labels.

\subsection{Anomaly Detection}

When a patch is learned for a label $z$, we apply it to the corresponding low-confidence data $D_{l}$, and calculate the transfer ratio $r$. Specifically, $r$ is defined as $ r = \frac{\sum_{x \in D_l} \mathbb{I}(g(x') \neq z)}{n}, \quad  x \in D_{l}$,
where $g(x)$ is the predicted label of $x$, and $\mathbb{I}$ is an indicator function. 
Based on the transfer ratio $r$ for each label, we use the MAD outlier detection method to compute the {\em anomaly index}.
According to the commonly-used three-sigma rule suggested by~\cite{iglewicz1993detect}, we set the threshold of anomaly index to 3.5, meaning that the current model is considered to be trojaned if its anomaly index is greater than this threshold.
Further, if all the transfer ratios are greater than 90\% for all labels, we consider the model to be clean even when the anomaly index is above the threshold. This could happen when, e.g., one transfer ratio is 95\% and the rest transfer ratios are all 100\%. 

\hide{It first calculates the absolute deviation between all data points and the median. The median of these absolute deviations is called MAD. Then the anomaly index of the data points are defined as the absolute deviation of the data points divided by the MAD. It assumes that the underlying distribution is a normal distribution so a constant estimator value (1.4826) is used to normalize the anomaly index. When the anomaly index is greater than a certain threshold, it means that the data point is an outlier.
}


{\em Remarks}. Our approach is similar to \nc~\cite{wang2019neural}, but bears several subtle and important differences. 
First, \name and \nc are built upon different observations. \nc assumes that the triggers are of small size while we focus on the prediction behaviors related to confidence. 
Second,  \name and \nc use different optimization objective functions. \nc learns a shortcut on clean data that leads to a specific label. In contrast, \name learns a patch on training data (can be both clean or poisoned) in a certain label and the patch leads the prediction to any other labels but the current one.
Third, \name and \nc compute anomalies on different metrics. \nc applies anomaly detection to the $L_1$ norms of the reversed triggers, while \name applies anomaly detection on the transfer ratios of low-confidence data.


\section{Evaluation}

\subsection{Experimental Setup}



{\bf Datasets}.
We use four datasets in our experiments including CIFAR10~\cite{krizhevsky2009learning}, GTSRB~\cite{stallkamp2011german}, ImageNet~\cite{deng2009imagenet}, and PubFig~\cite{kumar2009attribute}. All the datasets are publicly available. For GTSRB data, we use its subset with 13 labels provided by~\cite{liu2020reflection}. For PubFig and ImageNet, we randomly select a subset of them and each dataset contains 16 labels. We also train different neural network models for these datasets, and the details can be found in  Appendix~\ref{App:datasets}.

\begin{table}[t]
\centering
\caption{The evaluated backdoor attacks.}
\resizebox{1\columnwidth}{!}{
\begin{tabular}{ccrrr}
  \toprule
  \multirow{2}{*}{Attack} & \multirow{2}{*}{Dataset} & Original & Benign  & ASR (\%) \\
   & & Acc. (\%) & Acc. (\%) & \\
  \midrule
  \badnet & CIFAR10 & 85.85 & 85.05 & 96.34 \\
  \tronn & PubFig & 83.88 & 79.75 & 98.25 \\ 
  \cl & CIFAR10 & 85.85 & 84.06 & 96.12 \\
  \sig & ImageNet & 84.00 & 77.00 & 97.75 \\
  \refool & GTSRB & 99.32 & 98.07 & 99.98 \\
  \bottomrule
\end{tabular}
}
\label{tab:attack_info}
\end{table}

\noindent{\bf Backdoor Attacks}.
We evaluate five existing backdoor attacks, including \badnet~\cite{gu2017badnets}, \tronn~\cite{liu2018trojaning}, \cl~\cite{turner2018clean}, \sig~\cite{barni2019new}, \refool~\cite{liu2020reflection},
and \Ftrojan~\cite{wang2022invisible}.
The first two are change-label attacks and the rest three are clean-label attacks.
We choose these attacks as they are consistent with our threat model, i.e., launching attacks by poisoning the training data. The white-box attacks that need control of the model or the training process (e.g.,~\cite{nguyen2020input,nguyen2021wanet}) are out of our scope.
There are different combinations of attacks and datasets. We only apply \cl on CIFAR10 as the attack is ineffective on the other datasets, resulting in 17 attack combinations in total.
For the implementation, we use the open source code provided by the authors except for \cl.\footnote{The open source implementation of \cl is unavailable. We implement \cl ourselves and our implementation achieves almost the same results with the original paper.} 
We set the target label index to 0 for all the attacks by default.
For \badnet and \tronn, we define a square-shaped trigger with size 4 $\times$ 4 for $32 \times 32$ images and with size $32 \times 32$ for $224 \times 224$ images. These triggers are placed in the lower right corner of the image. For the other attacks, we use their default trigger setup.
To ensure an effective attack, we gradually increase the poisoning rate by 0.1\% each step until the ASR is above 95\%. The original accuracy of the model, the benign accuracy on clean inputs, and the ASR are shown in Table~\ref{tab:attack_info}, where we show the results of \badnet on CIFAR10, \tronn on PubFig, \cl on CIFAR10, \sig on ImageNet, and \refool on GTSRB  for brevity.
For advanced attacks, we implement them using \badnet on CIFAR10. In partial attack, we only poison label 1 data to the target label 0. 
In MTOT attack, we define three triggers at the lower-right, lower-left, and upper-left corners, respectively.
In MTMT attack, we use the same three triggers as in MTOT, but for target labels 0, 1, and 2, respectively. 

\noindent{\bf Backdoor Defenses}.
We compare with three state-of-the-art backdoor defenses that aim to detect whether a given model is trojaned, i.e., \nc~\cite{wang2019neural}, \abs~\cite{liu2019abs}, \ulp~\cite{kolouri2020universal}, and
\meta~\cite{xu2019detecting}.
For all of them, we use the open source code provided by the corresponding authors, and use their default setting.\footnote{Note that \abs and \ulp only provide implementations for the CIFAR10 dataset. That is, \abs provides executable binaries and \ulp provides the pre-trained meta-classifier for CIFAR10.}
For \name, there are two main parameters, i.e., $\lambda$ and $|D_h|$ in Eq.~\eqref{eq:reg}. For $\lambda$, we initialize it to 0.0001 and dynamically adjust it to a proper value following~\cite{wang2019neural}. For the sampling size, we set it to 50 by default, i.e., $|D_h|=|D_l|=50$. We will evaluate the performance of \name as this parameter varies.
The experiments are run on a machine with 20-cores Intel i9-10900KF CPU, 256GB RAM, and one NVIDIA GeForce RTX3090 GPU.\footnote{We will make the code publicly available upon acceptance.}

\hide{
\begin{figure*}[t]
\centering
\includegraphics[width=1.5\columnwidth]{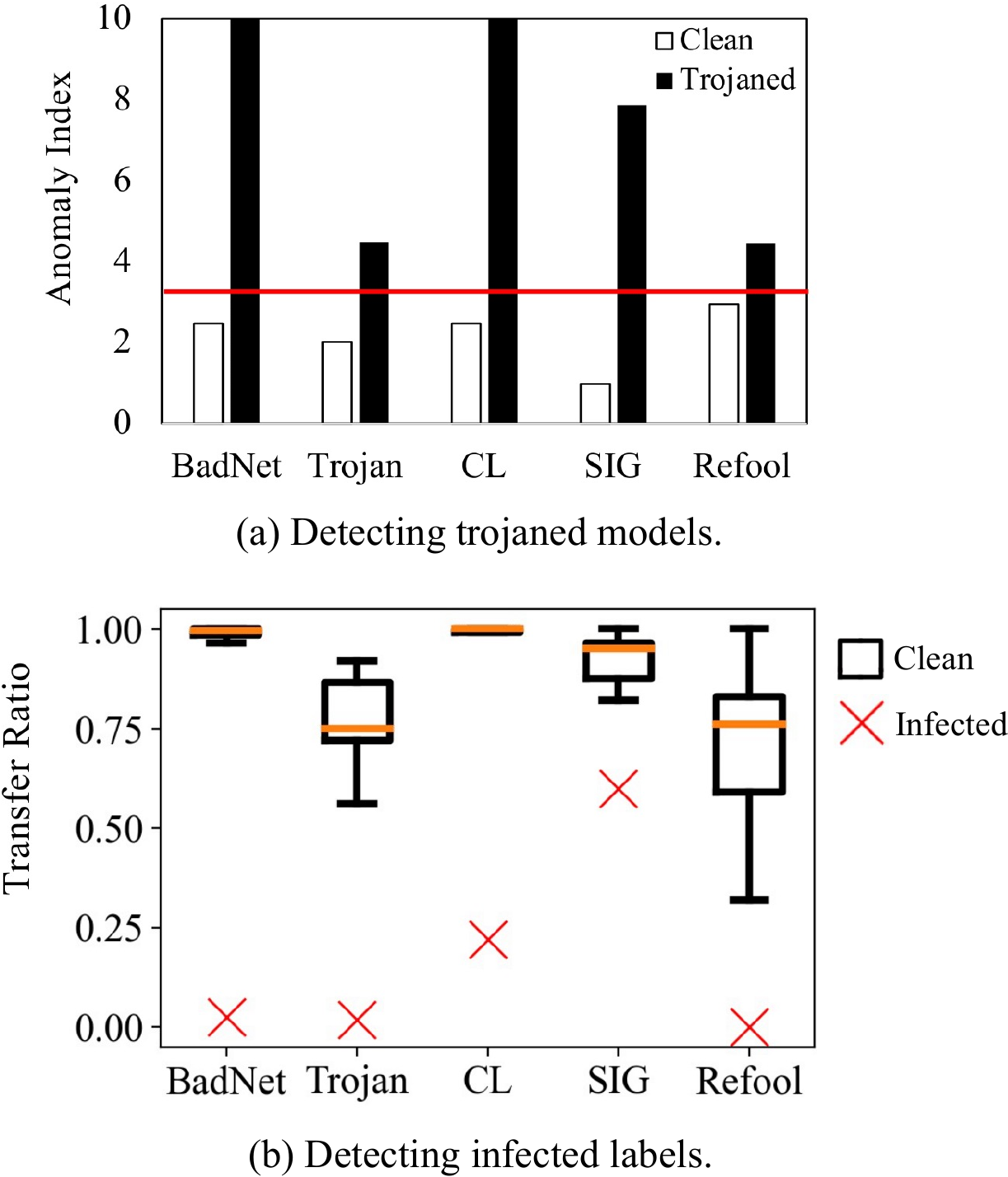}
\caption{The detection results against five existing backdoor attacks. The left and right figures show that \name can successfully detected trojaned models and the infected labels, respectively.}
\label{fig:fig_basic_exp}
\end{figure*}
}

\subsection{Experimental Results}\label{sec:exp1}


{\em (A) Detection against basic backdoor attacks}.
We first evaluate the detection accuracy of \name under the 17 combinations of basic backdoor attacks. The complete results are shown in Appendix~\ref{app:combinations}. In summary, \name can accurately distinguish the trojaned/clean models as well as identify the infected labels in 16 out of the 17 combinations.
Here, we further show the results of \badnet on CIFAR10, \tronn on PubFig, \cl on CIFAR10, \sig on ImageNet, and \refool on GTSRB in Figure~\ref{fig:fig_basic_exp}.
We can observe that the anomaly indices of trojaned models are significantly larger than that of clean models, and \name successfully distinguishes them.
Additionally, 
the infected label is identified as an outlier (denoted by red crosses), meaning that \name can also successfully indentify the infected label.

\begin{figure}[t]
\centering
\includegraphics[width=0.8\columnwidth]{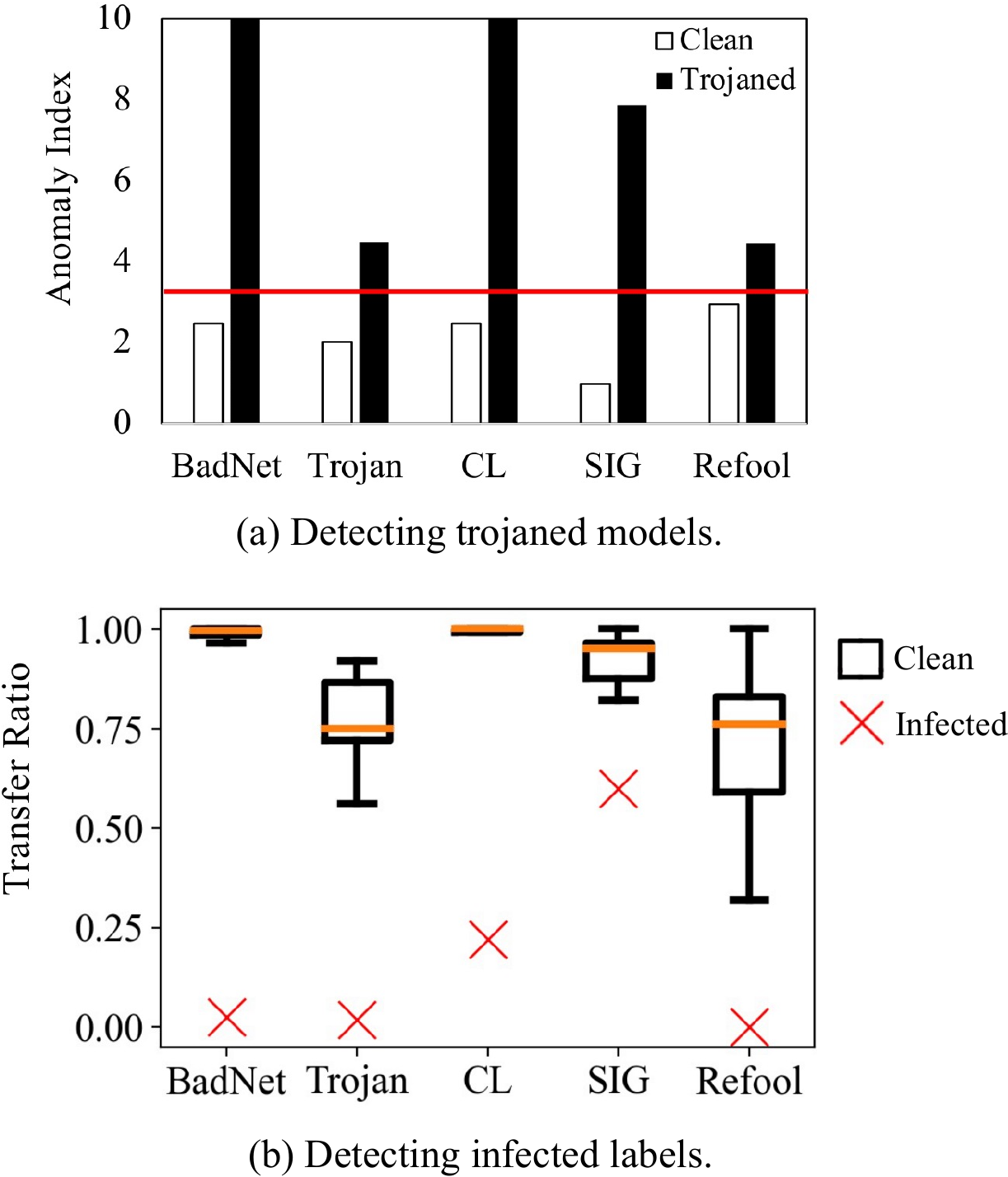}
\caption{The detection results against five existing backdoor attacks. \name can successfully detected trojaned models and the infected labels.}
\label{fig:fig_basic_exp}
\end{figure}

\begin{figure}[t]
\centering
\includegraphics[width=0.6\columnwidth]{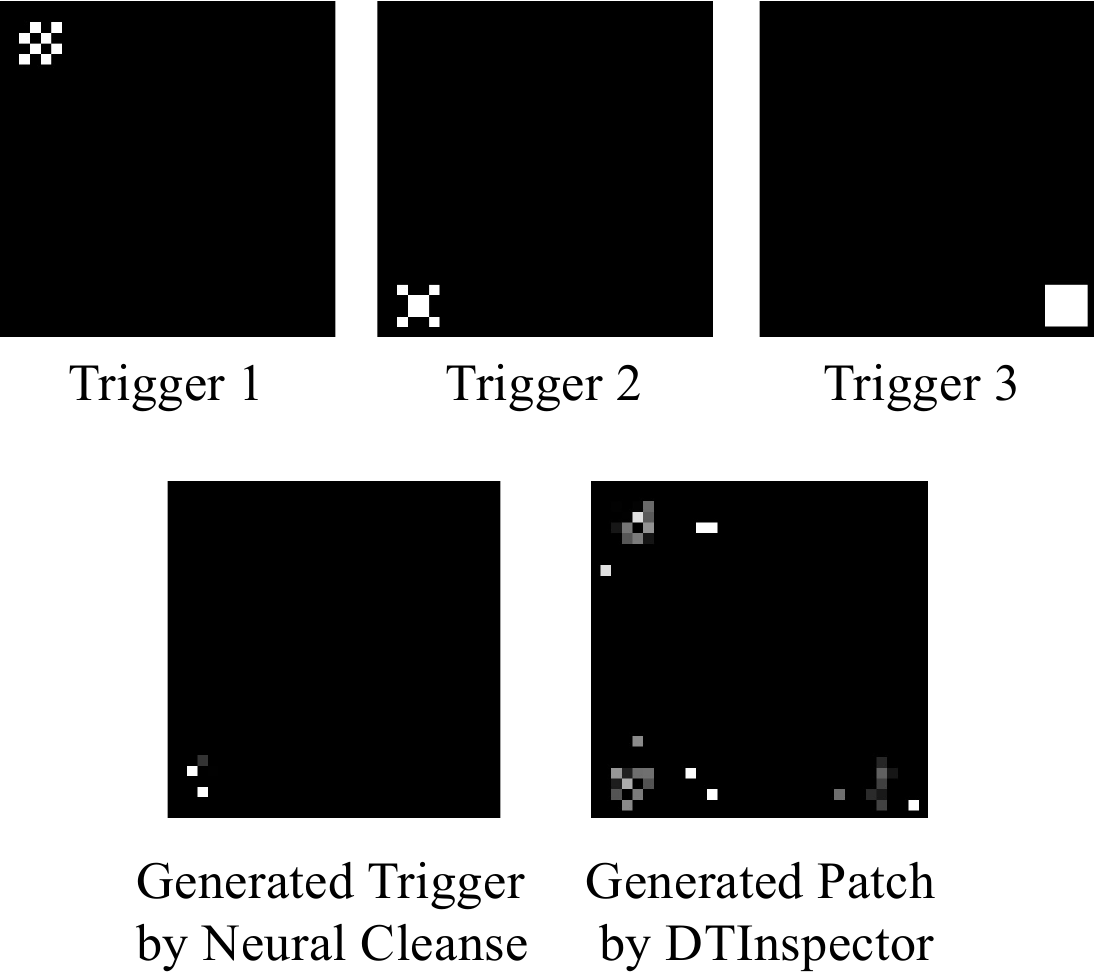}
\caption{Detected triggers by \nc and \name under the MTOT attack. \nc reverse-engineers one trigger while \name finds all the three triggers (patches).}
\label{fig:MTOT_Trigger}
\end{figure}

\begin{table}[t]
\centering
\caption{Detection results against advanced types of backdoor attacks. Compared to the competitors, \name successfully detect all such attacks.}
\begin{tabular}{crrrrr}
  \toprule
   Attack Type &\nc  & \abs & \ulp & \name \\
  \midrule
 Partial attack & Y & N & Y & Y (16.19) \\
 MTOT attack & Y & N & Y & Y (36.74) \\
 MTMT attack & N & N & N & Y (62.72) \\
  \bottomrule
\end{tabular}
\begin{tablenotes}
\small \item[1]* For all the tables, unless otherwise stated, `Y' means the model is identified as trojaned and `N' indicates otherwise; numbers in the parentheses are anomaly indices, and the model is considered to be trojaned if the index exceeds 3.5 for \name. \normalsize
\end{tablenotes}
\label{tab:advance}
\end{table}
{\em (B) Detection against advanced types of backdoor attacks}.
Next, we show the detection results against partial attack, MTOT attack, and MTMT attack.
The detection results of \name, \nc, \abs, and \ulp are shown in Table~\ref{tab:advance}. We can see that \name can detect these advanced backdoor attacks. In contrast, \abs fails to detect them (the REASR scores from both feature sapce and pixel space are zeros). This is probably due to the fact that \abs analyzes only one neuron and these attacks may involve multiple neurons. \ulp cannot detect the MTMT attack, and the possible reason is that the trained meta-classifier cannot generalize well to the complex model behavior after MTMT attack. \nc also cannot detect the MTMT attack. The reason is that \nc identifies multiple reversed triggers in this case, which tends to decrease the degree of abnormality. For the MTOT attack, we further show the detected triggers in Figure~\ref{fig:MTOT_Trigger}. We can see that although \nc can detect the attack, it identifies only one trigger; in contrast, \name can detect all the three triggers since we need to find all the triggers so as to flip the predictions.

\hide{
\yy{this is a small weakness of our method. we need to mention that we can find the infected label first.}
We also find that MTMT can only find one infected label, but cannot completely detect all target labels. That is to say, our detection method can detect the existence of backdoors for MTMT attacks, but there is no way to completely find all backdoors. This is because labels are divided into three types of labels in patch generation: uninfected labels (Labels A), infected labels with relative bigger trigger (Labels B) and the infected label with the smallest trigger (Label C). When the patch is generated, the patch shape of Labels A and Labels B is inclined to become the trigger shape of Label C under the L1 norm restrict. In this way, when both high-confidence data and low-confidence data are attached to this patch, the two transfer ratios will be high. However, the patch of Label C will destroy its trigger and make the high confidence data become benign data and keep the low confidence data still so that label C will be detected out. In summary, our Data Cleanse method can detect the existence of the above three kinds of attacks in the model. 
}

\begin{table}[t]
\centering
\caption{Comparisons with \ulp on unseen triggers. \name can detect the trojaned models while \ulp cannot.}
\begin{tabular}{crr}
  \toprule
  New Trigger & \ulp & \name \\
  \midrule
    8 $\times$ 8 colored square & N & Y (122.00) \\
    12 $\times$ 12 red square & N & Y \,~(39.23) \\ 

  \bottomrule
\end{tabular}
\label{tab:ULP}
\end{table}
{\em (C) Detection against unseen triggers}.
Next, we further compare \name with \ulp. As mentioned before, \ulp may not generalize well to unseen triggers. 
Here, we still use \badnet attack on CIFAR10, and define two triggers that are not included in the training data of \ulp, i.e., a randomly colored 8 $\times$ 8 square and a red 12 $\times$ 12 square. The detection results are shown in Table~\ref{tab:ULP}. We can see that \ulp cannot detect such attacks, which is consistent with our previous analysis. In contrast, \name can still detect such attacks.



\begin{table}[t]
\centering
\caption{Sensitivity comparisons with \nc on trigger size. \name is more robust to trigger size than \nc.}
\begin{tabular}{crrrr}
  \toprule
  \multirow{2}{*}{Trigger Size} & \multicolumn{2}{c}{Detection Result} \\
   \cmidrule(lr){2-3} & \nc & \name \\  
  \midrule
  4 $\times$ 4 & Y (2.62)  & Y (199.65) \\
  7 $\times$ 7 & Y (2.17)  & Y~~  (53.28) \\
  10 $\times$ 10 & Y (2.24)  & Y (121.03) \\
  13 $\times$ 13 & N (1.88)  & Y~~ (66.77) \\
  16 $\times$ 16 & N (1.36)  & Y~~  (36.51) \\
  19 $\times$ 19 & N (1.44)  & Y~~  (62.70) \\
  22 $\times$ 22 & N (1.27)  & Y~~~~ (5.39) \\
  \bottomrule
\end{tabular}
\label{tab:triggersize}
\end{table}

{\em (D) Sensitivity to trigger size}.
In Eq.~\eqref{eq:reg}, we use $L_1$ norm to restrict the size of the patch, which might cause our detection method to be sensitive to the trigger size. Here, we perform a sensitivity study by varying the trigger size. We apply \badnet on CIFAR10 and fix poisoning rate to 5\%. For the trigger, we gradually increase its size from $4 \times 4$ to $22 \times 22$. The results are shown in  Table~\ref{tab:triggersize}.
We can observe that \name is more robust to trigger size than \nc. \nc cannot detect the trojaned model when the trigger size is no smaller than $13 \times 13$, while \name can make a successful detection even when the trigger size grows to $22 \times 22$. Considering that the image size of CIFAR10 is $32 \times 32$, this means that \name can still work even nearly half of the image contains poisoning pixels.




{\em (E) Sensitivity to sampling size}.
One parameter of \name is the sampling size, i.e., the number of high- and low-confidence samples.
We still use \badnet attack on CIFAR10, and fix the trigger size to 4 $\times$ 4, and the poisoning rate to 5\%. We then vary the sampling size from 50 to 2000, and found that \name is relatively robust to the sampling size in a wide range (the detailed results are included in Appendix~\ref{app:sensitivity} for completeness). Specifically, it successfully detects the backdoor in a wide range, even when only 50 images are sampled from both high- and low-confidence data. This makes \name practically feasible as an effective backdoor attack usually needs to poison at least hundreds or thousands of images. 
\hide{When the sampling size is greater than 2000, the proposed method becomes less accurate as there will be more clean data in high confidence and it is difficult to generate a good patch as the detection result will fail.
In fact, the larger the number of samples, the more stable the generated patch, which can better destroy the area of the trigger, but it may make the high confidence data contain more clean data. Too few samples will make the optimization process unstable. (Falling into the condition like an adversarial attack that a small perturbation can only change a certain image's label.)}

\begin{table}[t]
\centering
\caption{Comparisons with data inspection method. The wrong detection results are underlined. Data inspection results in high mis-detection especially for the clean models.}
\resizebox{1\columnwidth}{!}{
\begin{tabular}{crrrr}
  \toprule
  \multirow{2}{*}{Attack} & \multicolumn{2}{c}{Data Inspection} & \multicolumn{2}{c}{\name}\\
   \cmidrule(lr){2-3}\cmidrule(lr){4-5} & Clean & Trojaned & Clean & Trojaned \\  
  \midrule
    \badnet & \underline{Y} ~~(6.95) & Y ~~(31.90) & N (2.47) & Y (199.65) \\
    \tronn & N ~~(1.88) & Y ~~(14.70) & N (2.02) & Y ~~~~(4.47) \\
    \cl & \underline{Y} ~~(6.95) & Y ~~(38.85) & N (2.47) & Y ~~(77.34) \\
    \sig & N ~~(1.76) & \underline{N} ~~~~(2.82) & N (0.98) & Y ~~~~(7.86) \\
    \refool & \underline{Y} (27.36) & Y ~~(13.47) & N (1.21)& Y ~~~~(6.34) \\
    
    Partial & \underline{Y} ~~(6.95) & Y (152.48) & N (2.47) & Y ~~(16.20) \\
    MTOT & \underline{Y} ~~(6.95) & Y ~~(21.42) & N (2.47) & Y ~~(36.74) \\
    MTMT & \underline{Y} ~~(6.95) & Y ~~(48.78) & N (2.47) & Y ~~(62.72) \\
  \bottomrule
\end{tabular}
}
\label{tab:direct_inspect}
\end{table}


{\em (F) Comparison with more related methods}.
Here, we compare \name with additional baselines that are adapted from existing work. Specifically, we first compare with a data inspection method whose goal is to spot poisoning training samples from clean one (see the related work section for more details). We still use \badnet attack on CIFAR10, and fix the trigger size to 4 $\times$ 4, and the poisoning rate to 5\%.
To setup the method, we sample the high-confidence and low-confidence data the same with \name, but calculate the KL divergence between their averaged representations from the model's penultimate layer (normalized by softmax) before applying the anomaly detection. 
The results are shown in Table~\ref{tab:direct_inspect}. We can see that the data inspection method has a high mis-detection rate especially for clean models. This is due to the fact that the trigger is usually of small size and thus difficult to distinguish. In contrast, we propose a distribution transfer technique to magnify the differences between clean data and poisoning data.

We also adapt an input filtering method STRIP~\cite{gao2019strip}, which is originally proposed to identify poisoning input at the inference stage. We use the default setting,\footnote{We also tune the two parameters of entropy boundary and clean data size, and found little difference.} and apply it on the training data. The results show that STRIP always mis-classifies a relatively large set of clean data as poisoned. This makes it less effective when the poisoning rate is low. For example, on the clean CIFAR10 data, STRIP identifies 163 poisoning samples. When we inject 250 poisoning samples with poisoning rate 0.5\% (ASR is larger than 90\%), STRIP identifies 225 poisoning samples, 65 out of which are clean data. 

\begin{table}[t]
\centering
\caption{Detection results against adaptive attacks. The attacker tries to reduce the prediction confidence via varying the poisoning rate. \name is still effective as long as the backdoor attack is effective (e.g., the ASR is above 70\%).}
\begin{tabular}{rrrr}
  \toprule
  Poisoning & Benign & \multirow{2}{*}{ASR (\%)} & \multirow{2}{*}{Detection Result} \\
   Rate (\%) & Acc. (\%) &  &  \\
  \midrule
  50.00 & 80.33 & 98.23 & Y ~~(51.94) \\ 
  40.00 & 81.24 & 97.80 & Y ~~(84.64) \\
  30.00 & 83.44 & 97.33 & Y ~~(30.35) \\
  20.00 & 83.94 & 97.35 & Y ~~(18.20) \\
  10.00 & 84.95 & 96.85 & Y ~~(33.05) \\
  3.00 & 85.05 & 96.34 & Y (199.65) \\
  0.10 & 85.39 & 77.71 & Y ~~(32.04) \\
  0.09 & 85.13 & 75.53 & Y ~~~~(6.67) \\
  0.08 & 85.46 & 73.64 & Y ~~~~(6.24) \\
  0.07 & 85.62 & 72.25 & Y ~~~~(4.17) \\
  0.06 & 85.65 & 66.37 & N ~~~~(0.00) \\
  0.05 & 85.65 & 57.35 & N ~~~~(0.00) \\
  \bottomrule
\end{tabular}
\label{tab:adaptive_defense}
\end{table}

{\em (G) Detection against adaptive attack}.
We also conduct an adaptive attack where the attacker can control the poisoning rate in a range to intentionally lower down the prediction confidence of poisoning samples.
We perform \badnet attack on CIFAR10 as an example, and gradually decrease the poisoning rate from 50\% to see the detection performance of \name.
The results are shown in Table~\ref{tab:adaptive_defense}.
First, we can see that when the poisoning rate is high (e.g., above 30\%), the benign accuracy begins to decrease, violating the stealthiness requirement of a backdoor attack. Still, \name is able to detect the attack with poisoning rate 50\%.
Second, when the poisoning rate drops below 0.06\%, \name cannot detect the existence of the attack. However, in such poisoning rates, the ASR also drops to 66.37\%, rendering a less effective backdoor attack.

\section{Related Work}

\subsection{Backdoor Attacks}
Starting from the seminal work~\citet{gu2017badnets,chen2017targeted}, 
various backdoor attacks have been proposed. One major line of such attacks is to improve the robustness against backdoor defenses~\cite{yao2019latent,saha2020hidden,shokri2020bypassing,lin2020composite,he2021raba}.
Additionally, some researchers propose dynamic attacks that generate different triggers for different inputs~\cite{salem2020dynamic,nguyen2020input}, clean-label attacks that do not change the labels of training data~\cite{turner2018clean,barni2019new,zhao2020clean,liu2020reflection,nguyen2021wanet}, physical attacks that use physical objects as triggers~\cite{wenger2021backdoor}, and data-free attacks that do not need the access to training data~\cite{liu2018trojaning,tang2020embarrassingly,costales2020live,pang2020tale}.

\subsection{Defenses against Backdoor Attacks}
Existing backdoor defenses can be divided into four categories, i.e., {\em model inspection}, {\em data inspection}, {\em input filtering}, and {\em backdoor removal}.


Model inspection methods aim to detect whether a given model is trojaned. \nc~\cite{wang2019neural} detects the existence of small perturbations that lead to the possible target label based on a set of clean data, and is further improved by several later defenses~\cite{chen2019deepinspect,guo2020tabor,shen2021backdoor}. However, these defenses generally require a large amount of clean data and are reported to perform poorly when the trigger size increases.
Several work proposes to analyze neuron activations~\cite{liu2019abs,ma2019nic}, but becomes less effective against advanced attacks such as MTOT attack and MTMT attack. Meta-classifiers trained with a number of existing clean models and trojaned models are also trained to distinguish trojaned/clean models~\cite{xu2019detecting,kolouri2020universal}. However, the trained meta-classifier may overfit the training data and degrade on unseen triggers.

Data inspection methods in the second category aim to distinguish poisoning training samples from clean ones given that the current model is trojaned~\cite{tran2018spectral,chen2019detecting,hayase2021spectre}. For example, \citet{tran2018spectral} find spectral representation is a good indicator to separate poisoning samples from clean ones. 

Defenses in the third category function at the inference stage, and they aim to detect whether an input has been corrupted or how to erase triggers in the input~\cite{cohen2019certified,udeshi2019model,gao2019strip,doan2020februus}.
For example, STRIP~\cite{gao2019strip} is based on the intuition that corrupted images usually make consistent predictions even when strong perturbations are added; Februus~\cite{doan2020februus} deletes influential regions in an image and restores the deleted regions via GAN.

In the fourth category, defenses aim to remove the backdoor in the models~\cite{liu2018fine,zhao2020bridging,li2021neural,wu2021adversarial,guan2022few}.
For example, Fine-pruning~\cite{liu2018fine} combines neuron pruning and fine-tuning for backdoor removal;  
NAD~\cite{li2021neural} uses a teacher network trained on clean data to remove the backdoor of a student network.

Our work falls into the first category, but can also be used to remove backdoors (see Appendix~\ref{App:mitigation}).
Defenses in the second and third categories are built upon the premise that the given model is trojaned. Although we adapt two baselines of them in our experiments, the results are not promising. How to better and effectively adapt them for detecting whether a model is trojaned still needs future exploration. 







\section{Conclusion}
In this paper, we have proposed a new backdoor defense \name. Our key insight is that an effective backdoor attack usually results in high prediction confidence on the poisoned training data, so as to ensure high ASR on poisoning inputs. We provide both theoretical and empirical evidence for this observation, and then propose a distribution transfer technique to distinguish trojaned models from clean ones, by using the shortcut nature of triggers.
Extensive experiments demonstrate that the proposed defense: 1) can accurately detect the trojaned model as well as the infected label, and 2) outperforms existing defenses in terms of robustness to trigger size and effectiveness against advanced attacks and unseen triggers.
In the future, we plan to extend our idea into natural language models and more computer vision tasks.



\bibliographystyle{ACM-Reference-Format}
\bibliography{backdoor}

\newpage
\appendix
\onecolumn

\section{Detailed Explanations for the Limitations of Existing Defenses}
Here, we add more discussion about the limitation of three state-of-the-art backdoor defenses that aim to detect whether a given model is trojaned. 

{\bf \nc~\cite{wang2019neural}}. The observation behind this defense is two-fold. First, a backdoor trigger essentially creates a shortcut from the original label to the target label. Second, the trigger is usually of small size. Therefore, \nc proposes to reverse-engineer a trigger for each label based on a set of clean data, and then apply outlier detection to all of the reversed triggers' $L_1$ norms to detect the potential trigger. A significantly smaller reversed trigger indicates the existence of a backdoor. However, it is observed that \nc becomes futile when the trigger size is relatively large~\cite{guo2020tabor}. Recent attacks (e.g., \refool~\cite{liu2020reflection}) have also successfully bypassed the detection of \nc by dispersing the trigger into a larger area.


{\bf \abs~\cite{liu2019abs}}. The key observation behind \abs is that a backdoor attack essentially compromises the inner neurons of deep neural networks, and a trigger usually correlates to a neuron stimulating which can lead to a prediction label regardless of the given inputs. With this observation, \abs proposes to analyze the activation pattern of each neuron to detect backdoor attacks. However, ABS analyzes one neuron each time and cannot deal with the attacks compromising a group of neurons, making it fall short against advanced attacks involving multiple triggers or multiple target labels~\cite{gao2020backdoor}. 

{\bf \ulp~\cite{kolouri2020universal}}. \ulp is based on a simple but effective observation that benign models and trojaned models exhibit different behavior patterns. Based on this observation, \ulp learns the so-called universal litmus patterns and trains a meta-classifier based on a number of clean and trojaned models. It then uses the meta-classifier to decide if a model is trojaned or not. However, although the trained meta-classifier shows some generalization ability, it may still overfit the training models and fail to generalize to unseen attacks or triggers.

\section{More Experimental Results}
Next, we provide more experimental results. 

\subsection{The datasets and the classifiers}\label{App:datasets}

We summarize the statistics of the used datasets in Table~\ref{tab:datasets_info}.  The CIFAR10 dataset contains 60,000 color images of size $32 \times 32$ in 10 different classes. We split it into 50,000 training images and 10,000 test images. We train a CNN with six convolutional layers and two fully-connected layers.
For the other three datasets, we resize them all to size $224 \times 224$. For GTSRB, we have 4,772 training images and 293 test images. For PubFig, we have 12,800 training images and 3,200 test images. For ImageNet, we have 20,567 training images and 800 test images. We train a ResNet50 model~\cite{he2016deep} for ImageNet, and a ResNet34 model for both GTSRB and PubFig.

\begin{table}[h]
\centering
\caption{Summary of datasets and classifiers.}
\begin{tabular}{crrrr}
  \toprule
  Dataset & Train/Test Image & Input Size & Label & Classifier \\
  \midrule
  CIFAR10 & 50,000/10,000 & 32 x 32 & 10 & 6 Conv + 2 Dense \\
  GTSRB & 4772/293 & 224 x 224 & 13 & ResNet34 \\
  PubFig & 12,800/3,200 & 224 x 224 & 16 & ResNet50\\
  ImageNet & 20,567/800 & 224 x 224 & 16 & ResNet50\\
  \bottomrule
\end{tabular}
\label{tab:datasets_info}
\end{table}

\subsection{More empirical results for our observation}\label{App:observation}

\begin{figure}[h]
\centering
 \subfigure[The \cl attack]{
    \includegraphics[width=1.5in]{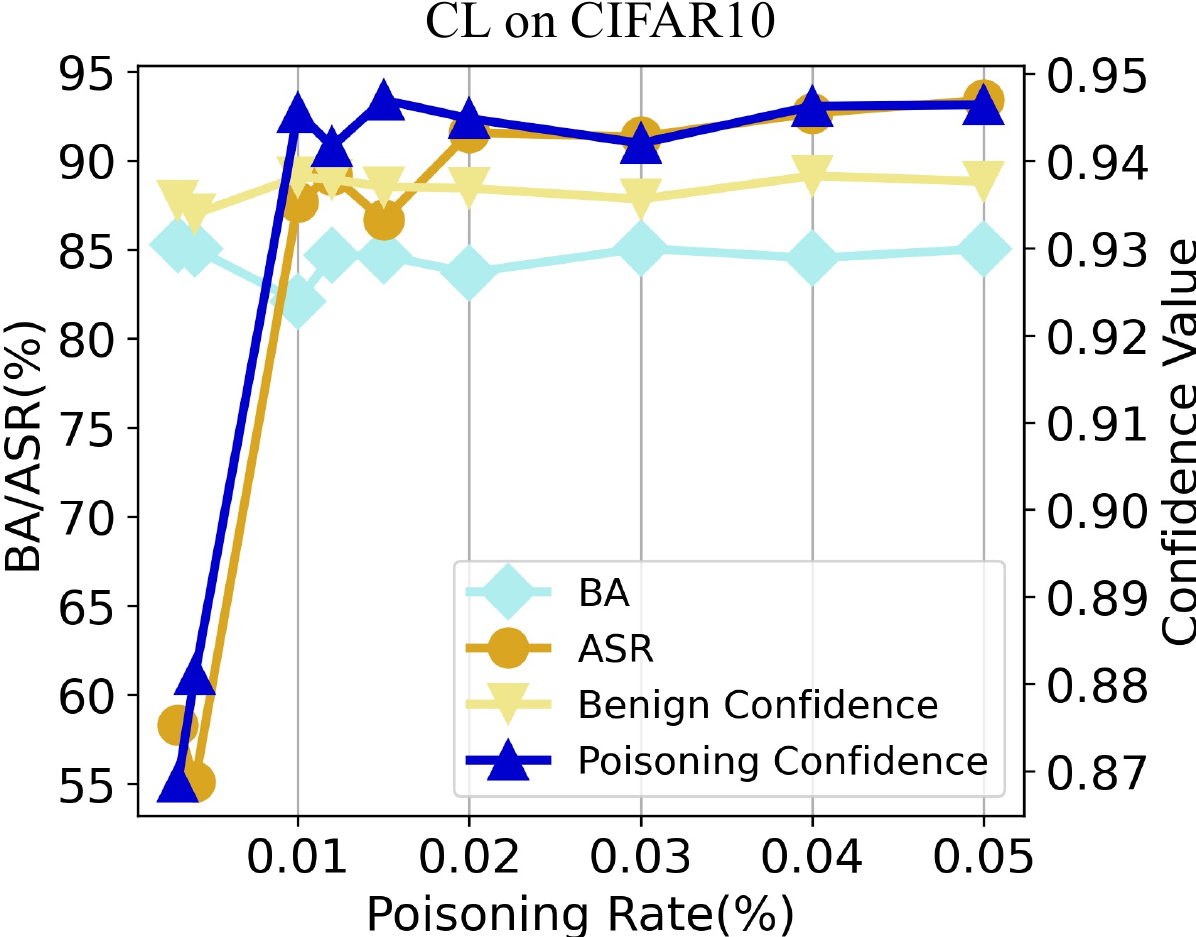}}
 \hspace{0.1in}
 \subfigure[The \tronn attack]{
    \includegraphics[width=1.5in]{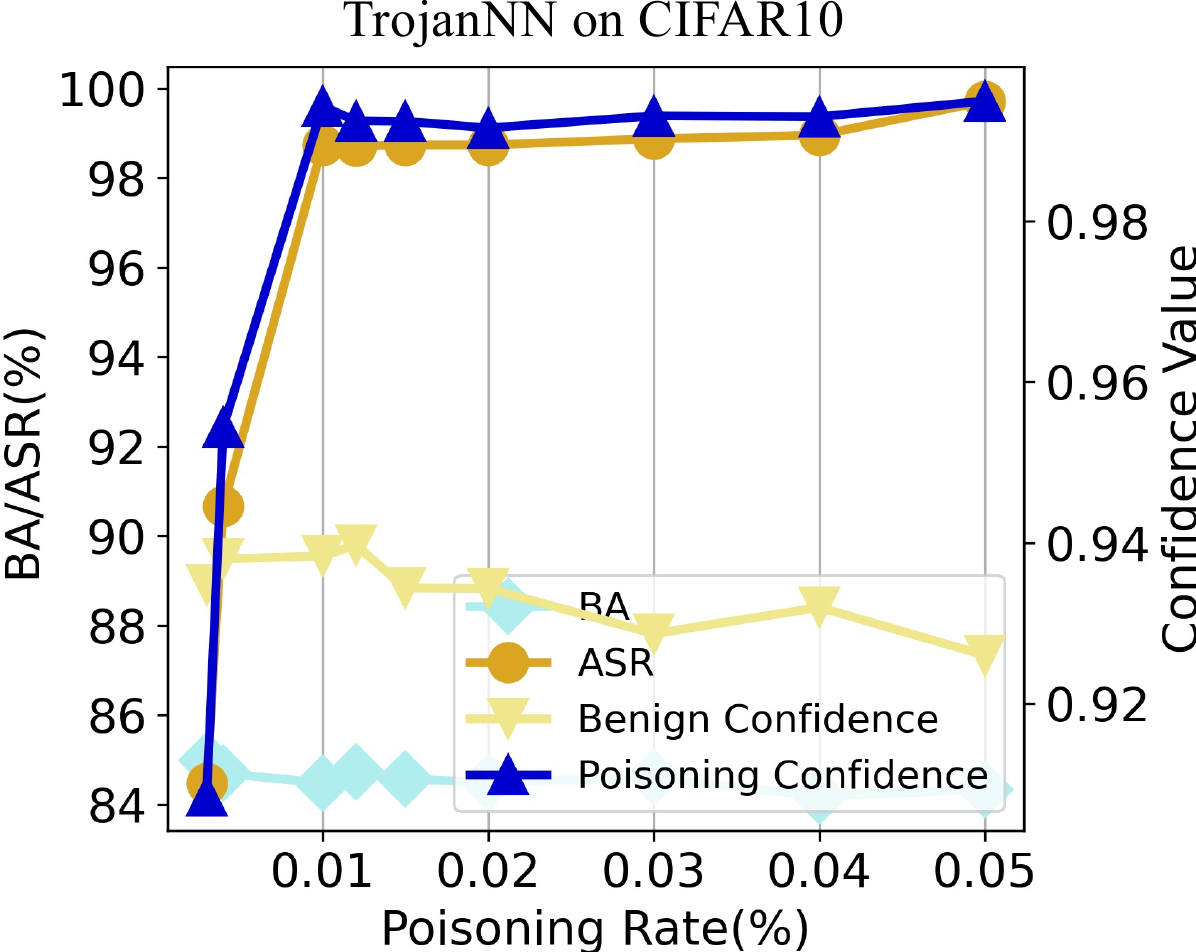}}
 \subfigure[The \sig attack]{
    \includegraphics[width=1.5in]{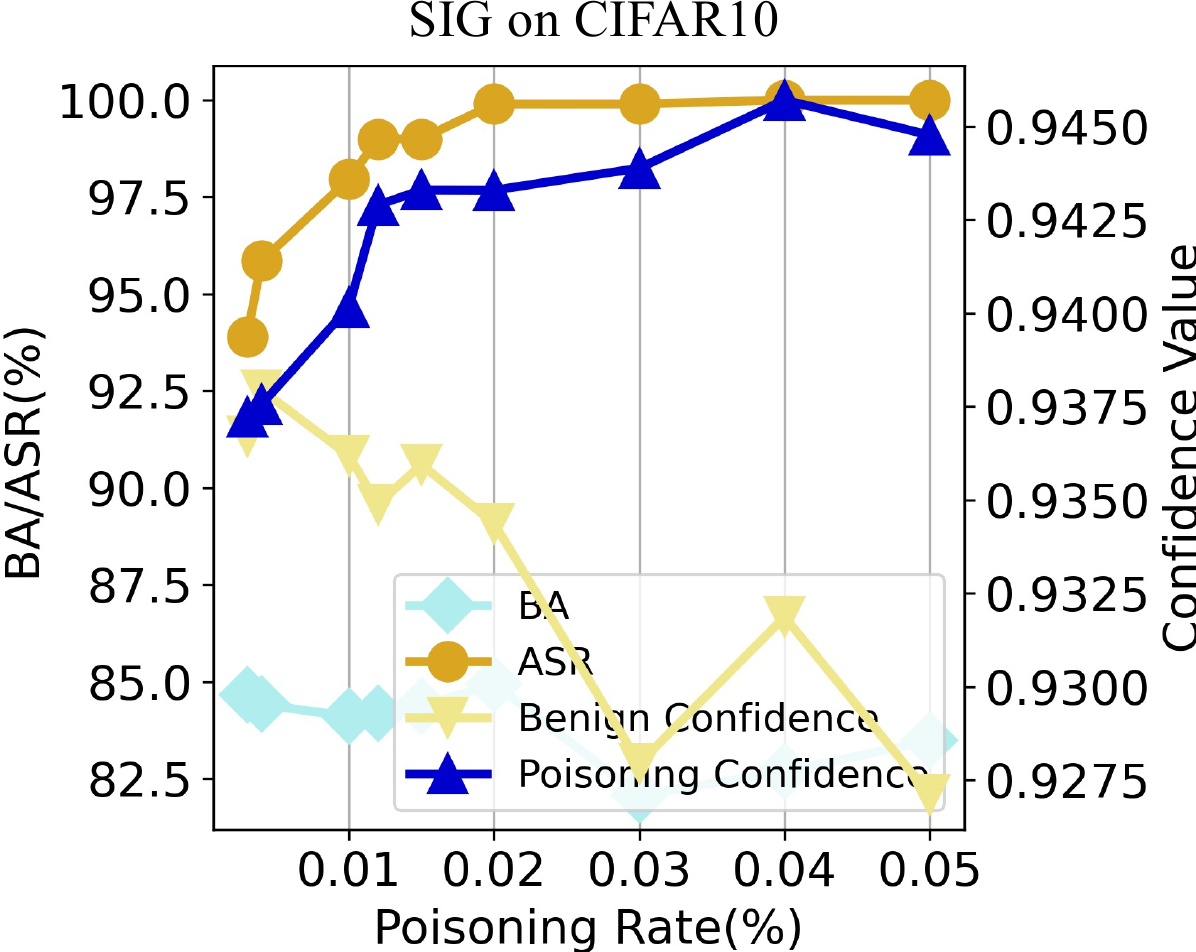}}
 \hspace{0.1in}
 \subfigure[The \refool attack]{
    \includegraphics[width=1.5in]{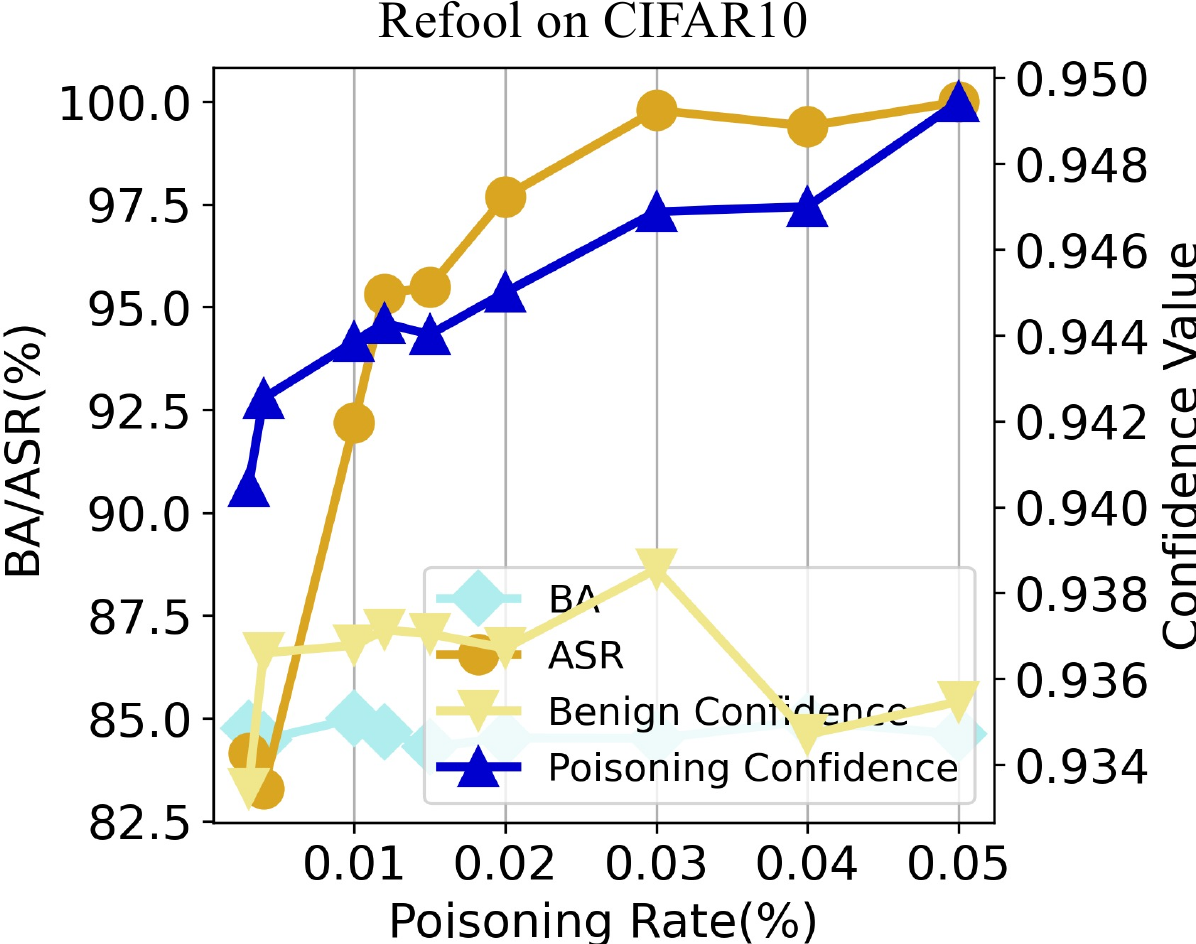}}
\caption{More empirical results for our observation. As the poisoning rate increases, the average prediction confidence of poisoning data becomes significantly higher than that of the clean data.}\label{fig:app_empirical}
\end{figure}

We first show the empirical results of more attacks for our observation. The result of \badnet is shown in Figure~\ref{fig:fig_distribution} and here we further show the results of \cl, \tronn, \sig, and \refool on CIFAR10 in Figure~\ref{fig:app_empirical}.  We can observe that the average prediction confidence of poisoning data is significantly higher than that of the clean data for all the four attacks.

\begin{table}[h!]
\centering
\caption{Detailed detection results of \name for all the attack and dataset combinations. The wrong detection is underlined.}
\begin{tabular}{ccrrr}
  \toprule
  Dataset & Attack &  Detection Result & Detect Infected Label\\
  \midrule
CIFAR10 & BadNet & Y(199.7) & Y\\
CIFAR10 & CL & Y(61.2) & Y\\
CIFAR10 & SIG & Y(13.5) & Y\\
CIFAR10 & Refool & Y(68.0) & Y\\
CIFAR10 & TrojanNN & Y(32.0) & Y\\
GTSRB & BadNet & Y(4.5) & Y\\
GTSRB & SIG & Y(3.66) & Y\\
GTSRB & Refool & Y(4.9) & Y\\
GTSRB & TrojanNN & Y(3.7) & Y\\
PubFig & BadNet & Y(7.1) & Y\\
PubFig & SIG & Y(12.1) & Y\\
PubFig & TrojanNN & Y(4.9) & Y\\
PubFig & Refool & \underline{N}(1.07) & \underline{N}\\
ImageNet & BadNet & Y(11.6) & Y\\
ImageNet & SIG & Y(7.7) & Y\\
ImageNet & Refool & Y(4.6) & Y\\
ImageNet & TrojanNN	& Y(8.4) & Y\\

  \bottomrule
\end{tabular}
\label{tab:all_result}
\end{table}

\subsection{Detection results for attack and dataset combinations}\label{app:combinations}
We next show the detailed detection results of \name under various attack and dataset combinations, and the results are shown in Table~\ref{tab:all_result}. We can observe that \name can successfully distinguish trojaned/clean models as well as the infected/clean labels in 16 out of the 17 combinations. The only exception is the \refool and PubFig combination. The reason is that the trigger of \refool covers the most influential areas (e.g., the central part) in the face images, and patching such areas may easily change the predictions. This limitation can be further addressed by, e.g., applying image restoration techniques such as GANs, and we leave this as future work. 

\begin{table}[t]
\centering
\caption{Sensitivity results on sampling size. \name works as this parameter varies in a wide range.}
\begin{tabular}{crrrr}
  \toprule
   Sampling Size & Detection Result \\
  \midrule
50 &  Y (199.6) \\
100 & Y ~~(31.4) \\
200 & Y ~~(65.4) \\
400 & Y (133.9) \\
800 & Y (103.6) \\
1500 & Y (327.0) \\
2000 & Y (166.8) \\
  \bottomrule
\end{tabular}
\label{tab:sampling}
\end{table}
\subsection{Sensitivity results of the sampling size}\label{app:sensitivity}


We include the detailed sensitivity results of the sampling size here. We still use \badnet on CIFAR10, and fix the trigger size to 4 $\times$ 4, and the poisoning rate to 5\%. We then vary the sampling size from 50 to 2000, and the results are shown in Table~\ref{tab:sampling}. 
We can see that \name is relatively robust to the sampling size in a wide range.
We also test different sampling methods to obtain the high-confidence and low-confidence samples (e.g., random sampling), and the results show little difference.

\begin{table*}[!t]
\centering
\caption{The mitigation results of \name. After removing the suspected data and retraining the model, the ASR can be reduced to less than 15\%, while the benign accuracy can be preserved in most cases.}
\begin{tabular}{crrrrrr}
    \toprule
    Attack & Original Acc. (\%) & Benign Acc. (\%) & ASR (\%) & Retrained Benign Acc. (\%) & Retrain ASR (\%) \\
    \midrule
    \badnet & 85.85 & 85.05 & 96.34 & 84.65 & 7.66 \\
    \tronn & 83.88 & 79.75 & 98.25 & 81.25 & 14.25 \\
    \cl & 85.85 & 84.06 & 96.12 & 84.12 & 8.93 \\
    \sig & 84.00 & 77.00 & 97.75 & 82.88 & 0.02 \\
    \refool & 99.32 & 98.07 & 99.98 & 97.92 & 0.11 \\
    Partial & 85.85 &  85.65 &  90.80 &  84.51 & 1.90 \\
    MTOT & 85.85 & 84.53 & 96.05 & 84.40 & 9.23 \\ 
    MTMT & 85.85 & 84.26 & 98.21 & 70.05 & 7.06 \\
    
    \bottomrule
\end{tabular}

\label{tab:mitigation}
\end{table*}

\subsection{Mitigation results using the proposed \name}\label{App:mitigation}
As mentioned before, the learned patch of \name can also be used to remove the backdoor.
Specifically, we consider the label whose transfer ratio is an outlier as the infected label.
For all the training samples belonging to this infected label, we superimpose the patch to each sample and remove the sample if it is classified to a label different from the original label. Otherwise, the sample is retained. We then retrain the model on the cleaned dataset. The results are shown in Table~\ref{tab:mitigation}. 

We can see that our mitigation method can mitigate most of the attacks (significantly lowering down the ASR), and meanwhile preserves the accuracy in most cases. One exception is from the MTMT attack, where the accuracy decreases with a relatively large margin. This is due to the fact MTMT attack infects multiple labels; therefore, we have to run the sample removing multiple times for MTMT attack, which significantly shrinks the training data size. This issue can be mitigated by, again, applying image restoration techniques to fix some of the poisoned images instead of directly deleting them, which is left as future work.

\section{Proof for Theorem 1}

Assume the data point $(X,Y)$ is sampled uniformly at random from $\cX \times \cY$, where $\cX = \bbR^d$ and $\cY = [c]$, i.e., there are $c$ labels. In the ideal case, the multi-class learning will optimize the following loss,
\begin{eqnarray*}
\cR(\bg) &=& \bbE_{(X,Y)\sim p(x,y)}[\ell(\bg(X), \be^Y)]\\
&=& \sum_{z=1}^c p(Y=z)\bbE_{X\sim p(x|Y = z)}[\ell(\bg(X), \be^z)]
\end{eqnarray*}
where $\bg: \cX\rightarrow \bbR^c$, $X\in \cX$, and $Y\in [c]$. $\be^Y$ is the standard canonical vector with the $Y$-th entry to be one and all others zero. $\ell$ is the defined loss function. We use cross-validation loss in our derivations.

By optimizing the risk, we could obtain a $\bg^*(\cdot)$, which is
\begin{eqnarray*}
\bg^* = \arg\max_{\bg\in \cG} \cR(\bg)
\end{eqnarray*}
with the prediction
\begin{eqnarray*}
\widehat Y = \arg\max_{i\in [c]} g_i(X)
\end{eqnarray*}
where $g_i(X)$ is the $i$-th entry of $\bg(X)$.

With the poisoning data by a backdoor attack, some data's label will be shifted to the target label $t$. In this way, the above ideal loss needs some modification. 
Without loss of generality, we denote the target label as $t$, and denote the poisoning rate in each label as $a_z$, i.e., $a_z = p(Y_a = t|Y=z)$ and $1 - a_z = p(Y_a = z|Y=z)$ where $Y_a$ is shifted to $t$ after the attack.
Then, we have
\begin{eqnarray*}
\cR_a(\bg) &=& \sum_{z} a_z p(Y=z)\bbE_{X\sim p(x|y = z)}[\ell(\bg(X), \be^{t})] + \\
&& \sum_{z} (1-a_z)p(Y=z)\bbE_{X\sim p(x|y = z)}[\ell(\bg(X), \be^{z})]
\end{eqnarray*}
whose minimizer is denoted as $\bg_a^*$. Here, we assume that the trigger involves very small perturbations and can be approximately ignored in the above equation.

In reality, we optimize an empirical version of $\cR(g)$, which is
\begin{eqnarray*}
\widehat \cR(\bg) = \sum_{i=1}^n \ell(\bg(x_i), \be^{y_i}).
\end{eqnarray*}

Next, we consider the term $\cR(\bg_a^*) - \cR(\bg^*)$. Note that
\begin{eqnarray}\label{E:tobebound}
\cR(\bg_a^*) - \cR(\bg^*) &\le& |\cR(\bg^*_a) - \widehat \cR(\bg^*_a)| + |\widehat \cR(\bg^*_a) -\widehat \cR_a(\bg^*_a)| \nonumber\\
& &+|\widehat \cR_a(\bg^*_a)- \widehat \cR(g^*) |  +|\widehat \cR(g^*) - \cR(g^*)|.
\end{eqnarray}

For the first and the fourth terms in the RHS of the above equation,
we could have the following two results assuming that the loss function $\ell$ is upper bounded by $M$,
\begin{eqnarray*}
|\cR(\bg^*_a) - \widehat \cR(\bg^*_a)| \le 2\fR_n(\ell\circ \cG) + M\sqrt{\frac{\log(2/\delta)}{2n}}\\
|\cR(\bg^*) - \widehat \cR(\bg^*)| \le 2\fR_n(\ell\circ \cG) + M\sqrt{\frac{\log(2/\delta)}{2n}}
\end{eqnarray*}
which could be easily proved by \emph{McDiarmid's inequality}~\cite{mcdiarmid} and the \emph{symmetrization}~\cite{DBLP:books/daglib/0097035}. Here $\fR_n(\cdot)$ denotes the Rademacher Complexity.
Note that if assuming $\ell(\bg(x), \be^Y)$ is Lipschitz continuous with a Lipschitz constant $L_\ell$, by the \emph{Talagrand's contradiction lemma}~\cite{DBLP:conf/alt/Maurer16} we have 
\begin{eqnarray*}\fR_n(\ell\circ \cG) \le \sqrt{2}L_\ell \sum_{y=1}^c \fR_n(\cG_y)
\end{eqnarray*}
where $\cG_y$ is the hypothesis space for $g_y(\cdot)$. 

For the second term in the RHS of Eq.~\eqref{E:tobebound}, we have
\begin{eqnarray*}
&&|\widehat \cR(\bg^*_a) -\widehat \cR_a(\bg^*_a)|\\
&=& \sum_{z} a_z \cdot p(Y=z)\sum_{x_i\in D_z}|\ell(\bg_a^*(x_i), \be^z) - \ell(\bg_a^*(x_i), \be^{t})|
\end{eqnarray*}
where we use $D_z$ to denote the subset of training data whose labels are changed to $t$ from label $z$. 

For the third term in the RHS of Eq.~\eqref{E:tobebound}, we have
\begin{eqnarray}\label{E:thirdterm}
&& |\widehat \cR(\bg^*) - \widehat \cR_a(\bg_a^*)| \nonumber\\
&=&\sum_{z} (1-a_z)p(Y=z)\sum_{x_i\in D_z}|\ell(\bg^*(x_i), \be^{z}) -\ell(\bg_a^*(x_i), \be^{z})| \nonumber\\
&=& +\sum_{z} a_z \cdot p(Y=z)\sum_{x_i\in D_z}|\ell(\bg^*(x_i), \be^z) - \ell(\bg_a^*(x_i), \be^{t})|.\nonumber\\
\end{eqnarray}
In the ideal case, backdoor attacks usually require that the predictions do not change for clean data, i.e., 
\begin{eqnarray*}
{g_a}_z^*(X) = g_z^*(X)
\end{eqnarray*}
for clean data (data points with both $Y_a = z$ and $Y = z$). 
The first term in the RHS of Eq.~\eqref{E:thirdterm} can be approximated to zero, i.e., $\ell(\bg^*(x_i), \be^{z}) \sim \ell(\bg_a^*(x_i), \be^{z})$.

Combining together, we have
\begin{eqnarray*}
\cR(\bg_a^*) - \cR(\bg^*) &\le& 4\fR_n(\ell\circ \cG) + 2M\sqrt{\frac{\log(2/\delta)}{2n}}+\\
& & 2 \sum_z a_z \sum_{x_i\in D_z}|\ell(\bg^*(x_i), \be^z) - \ell(\bg_a^*(x_i), \be^{t})|,
\end{eqnarray*}
which completes the proof.





\end{document}